\documentclass[prl,superscriptaddress,showpacs,preprintnumbers,showkeys,amsmath,amssymb,preprint]{revtex4}

\usepackage[dvips]{graphicx}
\usepackage{dcolumn}
\usepackage{bm}

\begin{document}
 
\title{Measurement of single electron emission in two-phase xenon}
\author{B.Edwards}
\email{blair.edwards@imperial.ac.uk}
\author{H.M.Ara\'ujo}
\affiliation{Blackett Laboratory, Imperial College London, UK}
\affiliation{Particle Physics Dept., Rutherford Appleton Laboratory, UK}
\author{V.Chepel}
\affiliation{LIP-Coimbra \& Department of Physics of the University of Coimbra, Portugal}
\author{D.Cline}
\affiliation{Department of Physics \& Astronomy, University of California, Los Angeles, USA}
\author{T.Durkin}
\affiliation{Particle Physics Dept., Rutherford Appleton Laboratory, UK}
\author{J.Gao}
\affiliation{Department of Physics, Texas A\&M University, USA}
\author{C.Ghag}
\author{E.V.Korolkova}
\affiliation{School of Physics, University of Edinburgh, UK}
\author{V.N.Lebedenko}
\affiliation{Blackett Laboratory, Imperial College London, UK}
\author{A.Lindote}
\author{M.I.Lopes}
\affiliation{LIP-Coimbra \& Department of Physics of the University of Coimbra, Portugal}
\author{R.L\"{u}scher}
\affiliation{Particle Physics Dept., Rutherford Appleton Laboratory, UK}
\author{A.St.J.Murphy}
\affiliation{School of Physics, University of Edinburgh, UK}
\author{F.Neves}
\affiliation{LIP-Coimbra \& Department of Physics of the University of Coimbra, Portugal}
\author{W.Ooi}
\affiliation{Department of Physics \& Astronomy, University of California, Los Angeles, USA}
\author{J.Pinto da Cunha}
\affiliation{LIP-Coimbra \& Department of Physics of the University of Coimbra, Portugal}
\author{R.M.Preece}
\affiliation{Particle Physics Dept., Rutherford Appleton Laboratory, UK}
\author{G.Salinas}
\affiliation{Department of Physics, Texas A\&M University, USA}
\author{C.Silva}
\author{V.N.Solovov}
\affiliation{LIP-Coimbra \& Department of Physics of the University of Coimbra, Portugal}
\author{N.J.T.Smith}
\affiliation{Particle Physics Dept., Rutherford Appleton Laboratory, UK}
\author{P.F.Smith}
\affiliation{Particle Physics Dept., Rutherford Appleton Laboratory, UK}
\affiliation{Department of Physics \& Astronomy, University of California, Los Angeles, USA}
\author{T.J.Sumner}
\author{C.Thorne}
\author{R.J.Walker}
\affiliation{Blackett Laboratory, Imperial College London, UK}
\author{H.Wang}
\affiliation{Department of Physics \& Astronomy, University of California, Los Angeles, USA}
\author{J.T.White}
\affiliation{Department of Physics, Texas A\&M University, USA}
\author{F.L.H.Wolfs}
\affiliation{Department of Physics and Astronomy, University of Rochester, New York, USA}

\date{\today}

\begin{abstract}
We present the first measurements of the electroluminescence response to the emission of single electrons in a two-phase noble gas detector. Single ionization electrons generated in liquid xenon are detected in a thin gas layer during the 31-day background run of the ZEPLIN-II experiment, a two-phase xenon detector for WIMP dark matter searches. Both the pressure dependence and magnitude of the single-electron response are in agreement with previous measurements of electroluminescence yield in xenon. We discuss different photoionization processes as possible cause for the sample of single electrons studied in this work.  This observation may have implications for the design and operation of future large-scale two-phase systems.
\end{abstract}

\maketitle

Early work investigating two-phase emission of ionization electrons was carried out in the 1940's \cite{hutchinson}, but the mechanism was not fully exploited as a method for radiation detection until the 1970s with the development of detectors using condensed argon \cite{dolgoshein}.  The field has since expanded, with the two-phase technique now being applied to WIMP dark matter searches, with coherent neutrino scattering and double $\beta$-decay experiments proposed \cite{darkmatter,hagmann}.  Some of these applications require detection of very small signals from rare events, demanding high sensitivity from the technique.

In the case of two-phase xenon, the interaction between an incident particle and a liquid xenon target produces prompt scintillation photons \cite{kubota,hitachi} in the vacuum ultra-violet (VUV) with $\lambda \simeq 175$~nm (FWHM $\simeq$ 14~nm) \cite{morikawa,hitachi_lambda}.  With no external electric field applied, ionization electrons created by the interaction will recombine, increasing the scintillation signal.  By applying an electric field across the liquid, some electrons can be extracted from the interaction site, to be detected independently.  The currently favoured method of low level charge detection from a liquid target relies on using the electroluminescence process to convert the ionization signal into a proportional photon yield in the gas phase.  Upon reaching the liquid surface, the ionization electrons must be emitted into the gas, a process dependent upon the electric field perpendicular to the surface \cite{Boloz1}.  Measurements of the emission coefficient in xenon, i.e. the fraction of electrons emitted into the gas, show that extraction approaches unity at 5 kV/cm \cite{gushchin}.  Once in the gas phase, the electrons are accelerated by a stronger electric field, exciting the gas atoms through collisions and causing the production of secondary VUV scintillation photons.  This electroluminescence process has been simulated and measured previously \cite{conde,santos,akimov,coimbra,aprile,sec_scint}, although some disagreement still remains over the absolute number of photons produced per electron.

Presented here is the first study quantifying the response to the emission of a single ionization electron in a two-phase noble gas detector.  In two-phase argon, a triple-GEM structure used for charge amplification in the gas has recently been reported to achieve single-electron sensitivity, although this study relied on electrons photo-produced in the first GEM rather than emitted from the liquid phase \cite{se_argon}.

In this article the single electron response of a two-phase xenon detector is described and the origin of the ionization electrons considered.  Sensitivity to single ionization electrons is important in experiments searching for very small, rare events.  On a technical level, it allows for a direct measurement of the ionization yields of different interacting particles, such as nuclear and electron recoils, and may help with the study of photoionization processes in liquid xenon.  Understanding the origin of these electrons may highlight new backgrounds for experiments relying on the detection of even smaller ionization signatures than those considered of interest in WIMP dark matter searches, which motivated this particular study.  One such example is the proposed detection of keV energy deposits from coherent neutrino scattering \cite{hagmann}.

\begin{figure}[!h]
\includegraphics[width=0.48\textwidth,clip]{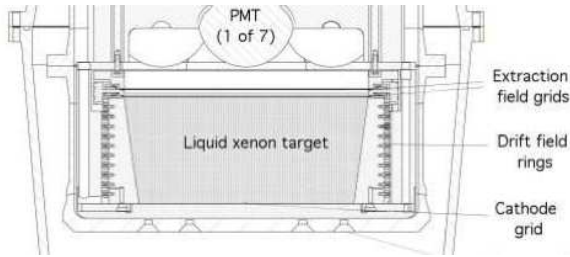}
\caption{\label{target} Schematic of the ZEPLIN-II detector.  The liquid and gaseous xenon regions are shown, along with the electric field electrodes and the array of PMTs detecting VUV photons.}
\end{figure}

This work was carried out using ZEPLIN-II, a two-phase xenon detector searching for WIMP dark matter.  The detector uses a 31 kg liquid xenon (LXe) target, held in a truncated cone of reflective PTFE, 140~mm deep with upper and lower diameters of 324 mm and 290 mm (Fig. \ref{target}).  The liquid surface lies between two meshes separated by 10~mm, where a strong electric field can be applied to produce electroluminescence from emitted electrons.  The details of the detector design, operation and primary analysis are detailed in Ref. \cite{sciencepaper}.  The extraction electrodes consist of a woven stainless steel mesh of 30~$\mu$m diameter wire, with a centre-to-centre separation of 500~$\mu$m.  The electrons are drifted towards the extraction region by a vertical 1~kV/cm electric field; upon reaching the liquid surface, they are emitted into the gas by an electric field of  $\sim$4.8~kV/cm and accelerated across the 2-3~mm gas gap by a field $\simeq$ 9.5 kV/cm.  Both the primary (scintillation) and secondary (electroluminescence) signals are  independently detected by the same array of seven photo-multiplier tubes (PMTs).  The time taken for the electrons to drift through the liquid provides separation in time proportional to the depth in the detector.

In this work data from the 31-day shielded run of ZEPLIN-II was used to search for evidence of single electron emission into the gas phase. During this run the average background rate from gamma-ray interactions was  $\sim$2 events/s ($>$5 keV). An unexpected population of very small secondary-like signals immediately following large secondaries was apparent during early tests. Some of these signals were also observed {\em between} the primary and secondary pulses associated with normal events and these were selected for further analysis.

The response of the detector to a single electron emitted from the liquid is predicted to be small, yielding fewer than 10 photoelectrons across the array of seven PMTs.  To search for these small signals, quiet timelines free from spurious noise, overlapping events and optical feedback effects are required.  For this reason the search for candidate single electron signals was carried out on the high-quality, low-background dataset used for WIMP searches and analysed in Ref. \cite{sciencepaper}.  From this dataset, only events triggered by the primary scintillation signal were selected in order not to bias the single electron signal distribution.  The trigger function is a five-fold coincidence at the level of a third of a single PMT photoelectron (phe).

\begin{figure}[!h]
\includegraphics[width=0.42\textwidth,clip]{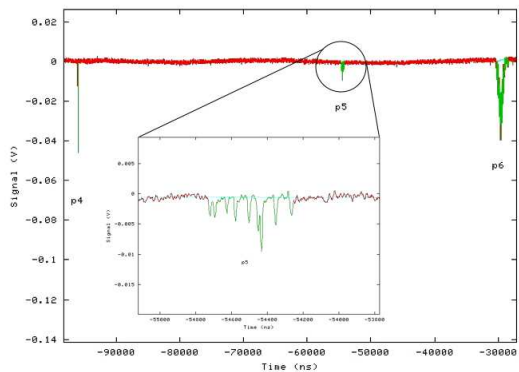}
\caption{\label{se_trace} Seven-PMT sum waveform containing a candidate single electron event.  The primary scintillation pulse (p4) and secondary electroluminescence pulse (p6) are clearly visible, with the single electron (p5) occuring in the quiet section of the timeline between the two.}
\end{figure}

Fig. \ref{se_trace} shows an example of an 80 $\mu$s waveform (sampled at 2 ns) containing a primary scintillation pulse and a secondary electroluminescence pulse, with a single electron candidate extracted in the intervening time.  The single electron pulse is detected as a collection of individual PMT photoelectrons spread over a $\sim$500~ns period, the time taken for the electron to cross the high-field electroluminescence region.

In the analysis described in Ref. \cite{sciencepaper}, the data were corrected for the finite electron lifetime in the liquid as well as operational parameters which affect the gain of the ionization channel (such as variations of pressure, liquid level and electric fields). In this analysis the `purity' correction, compensating the secondary scintillation signal for trapping of electrons by electronegative impurities during their drift to the liquid surface, is not required as a single electron will either reach the surface or be trapped, meaning no $partial$ loss of signal.  The other operational parameters are considered as variables.

\begin{figure}[!h]
\includegraphics[width=0.4\textwidth,clip]{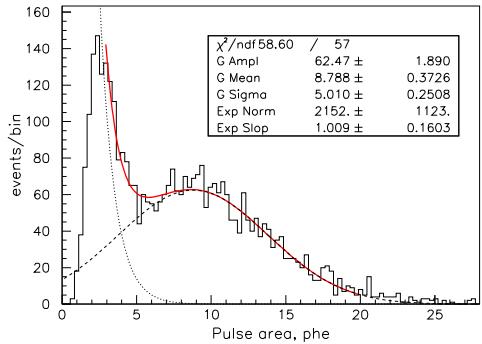}
\caption{\label{se_spectra} Example of a single electron spectrum, with gas pressure of  1.5 bar.  The continuous line shows the fit to the entire spectrum; the Gaussian and exponential  components are also shown.  The inset shows the gaussian plus exponential fit parameters.}
\end{figure}

A pulse area histogram of all small signals detected between primary and secondary pulses, representing the number of photoelectrons detected by the PMTs, is shown in Fig. \ref{se_spectra}.  The mean area of a single phe pulse for each PMT is independently measured.  The figure shows a clear population which we attribute to single electrons, along with an exponential noise pedestal.  The distribution of single electrons is fitted with a gaussian function.  Although phe statistics suggest a Poisson distribution, further broadening occurs due to electronic noise and other fluctuations.  The spectrum shows the single electron mean of 8.8$\pm$0.4~phe and a width $\sigma$=5.0~phe.  As discussed later this corresponds to over 200 VUV photons, demonstrating the large gain of the ionization channel.

The number of electroluminescence photons created by one electron depends on the pressure, electric field and gas thickness.  The electroluminescence yield of xenon, defined as the number of VUV photons produced per electron per cm travelled, has been studied mainly for gas at room temperature  (see \cite{sec_scint} and references therein).  A dependence is found of the form $Y=aE-bP_{eq}$, where \emph{a} and \emph{b} are experimentally determined parameters, \emph{E} is the electric field in the gas and \emph{$P_{eq}$} is the equivalent pressure for the same gas density at $0\,^{\circ}$C.  It is known that the photon yield in the cold, saturated vapour is higher than that in the warm gas (for the same density).  This effect is clearly shown in Ref. \cite{coimbra}, where the room temperature measurement is consistent with other published results, but the yield in the vapour is clearly higher.  We note that for the mesh design and thickness of the gas layer mentioned previously, a parallel and uniform electric field can be assumed without significant error.  This permits direct calculation of the absolute electroluminescence yield in ZEPLIN-II.  If the small signals under scrutiny correspond to the emission of a single electron, this yield must agree with that measured for the cold vapour and show the same pressure dependence.

\begin{figure}[!h]
\includegraphics[width=0.42\textwidth,clip]{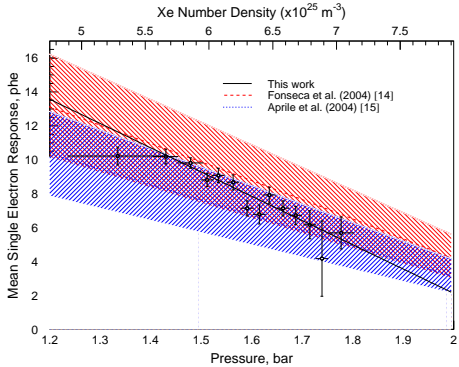}
\caption{\label{pressure} Dependence of size of single electron response on detector pressure, compared with predictions from previous measurements of electroluminescence yield in saturated vapour.  The shaded bands represent uncertainties assigned to these predictions, including an ad-hoc 10\% error assumed for parameters $a$ and $b$ (not given in the literature).}
\end{figure}

Gas pressure, $P$, is one of the main operational parameters affecting the size of the single electron response in the detector, doing so in two complementary ways: it affects electroluminescence at a microscopic level and is also directly linked to the liquid level through thermal expansion of the liquid, which is in thermal equilibrium with the gas.  The thermal expansion causes variation in the electric field in the gas, although this contribution is small.  Previous yield measurements can be compared directly with the results from ZEPLIN-II by factoring in the thickness of the gas region, $d$, the light collection efficiency, $\eta$=0.24, and the quantum efficiency, $QE$=0.17, of the PMTs.  The gas thickness is calculated from the drift time of background interactions occurring very near the lower extraction grid just under the liquid surface; the light collection was simulated by Monte Carlo; there is some uncertainty about the variation of PMT QE down to low temperatures for this particular phototube model, which we believe to be small.  Fig. 4 shows the mean single electron response (in phe), $Q\!E\,\eta\,d(P)\,Y(P)$, demonstrating the expected inverse dependence with xenon pressure over the range 1.2--1.9 bar, the variation observed during the science data run.  Both the absolute yield of VUV photons and the dependence on pressure are in good agreement with predictions made from the previous measurements of electroluminescence yield, providing strong evidence that the population observed is indeed from single electrons.

\begin{figure}[!h]
\includegraphics[width=0.4\textwidth,clip]{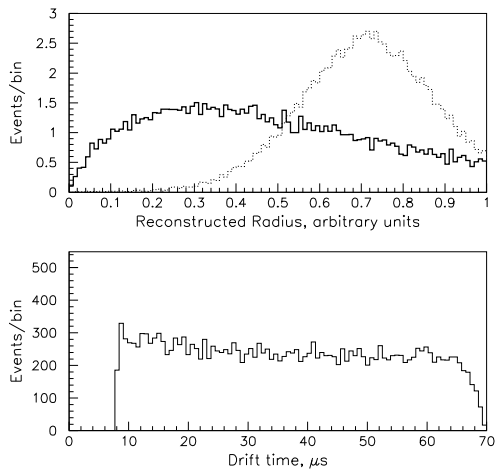}
\caption{\label{se_radial} Radial (top) and depth (bottom) distributions of single electrons measured.  In radial plot the solid line shows the single electron population, with the dashed line showing the distribution of small ionization signals ($\sim$few electrons) from detector walls \cite{sciencepaper}.}
\end{figure}

In ZEPLIN-II position reconstruction in the horizontal plane uses a simple centroid method, which naturally lacks precision for small signals such as those from single electrons.  However, the reconstructed radial distribution (shown in Fig. \ref{se_radial}) suggests an origin distributed over all radii which is clearly incompatible with that of small ionization signals originating from the detector walls (corresponding to a radius of 0.7 a.u.).  The later is a known background of nuclear recoils arising from plating of the walls with alpha-emitters from radon decay in the Xe \cite{sciencepaper}.  Furthermore the depth (electron drift time) distribution demonstrates relatively uniform production throughout the whole liquid xenon bulk.

\begin{figure}[!h]
\includegraphics[width=0.45\textwidth,clip]{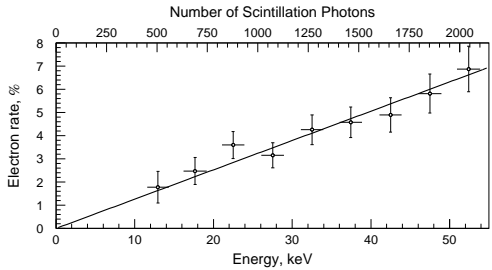}
\caption{\label{se_s1rates} Relative rates of single electrons as a function of the size of the primary signal (normalized to gamma-ray energies).  The rate is given as the percentage of timelines checked which contain single electron signals.}
\end{figure}

The fact that large secondary signals are seen to be followed by multiple single electron pulses suggests their production may be related to the number of VUV photons present in the chamber.  The quiet timelines found between primary and secondary signals, together with the proportionality between the primary signal and the energy deposited, allow testing of this hypothesis in a quantitative manner.  Fig. \ref{se_s1rates} shows the fraction of events where a single electron is observed as a function of energy (proportional to the number of VUV scintillation photons generated in the liquid).  A clear dependence on energy is observed, suggesting that the production of single electrons could be, at least in part, due to photoionization processes in the liquid bulk. 

The mean free path (mfp) of $\simeq$7.1~eV ($\lambda \simeq 175$~nm) scintillation photons for generating an electron in the liquid can be estimated from the event frequency shown in Fig. \ref{se_s1rates}. A scintillation yield of 39 photons/keV at 1 kV/cm has been measured, indicating an average 39,000 scintillation photons are required to produce an electron.  Photons generated in the liquid can escape from the surface or be absorbed in the surrounding PTFE reflectors or in the electrodes. A Monte Carlo simulation places the mean escape length in the liquid xenon at $\sim$25 cm for photons generated uniformly throughout the liquid. The latter value is not significantly affected by bulk absorption for the xenon purity considered here, or by Rayleigh scattering. This places the photon mfp for photoionization at $\sim$1.0x$10^{6}$ cm.

Several chemical species can be responsible, mainly related to impurites present in the liquid xenon. Unfortunately, the mfp depends on both the microscopic cross-section and the atom concentration, and so none can be ruled out with certainty since either (or both) quantities may be unknown.

Sub-threshold photoionization, either of impurities or of xenon atoms, dimers and higher order polymers, cannot be ruled out, even if they are unlikely. Some of the most abundant impurities are electronegative species (such as O$_{2}$, H$_{2}$O, N$_{2}$O, etc) responsible for the finite electron lifetime, and their concentration can be estimated from the rate of electron attachment. From known attachment cross-sections \cite{bakale} and considering an average energy of 0.1 eV for electrons drifting in a 1 kV/cm electric field \cite{gullikson}, we estimate that an O$_{2}$ concentration of $\sim$60 ppb or a N$_{2}$O concentration of just $\sim$8 ppb would produce the $\sim$100 $\mu$s electron lifetime observed during this ZEPLIN-II run. The photoionization cross-section needed to explain the measured mfp is only $\sim$0.001 Mb (compared to 10-100 Mb typically found well above threshold). In the case of intrinsic photoionization (the threshold in liquid xenon is 9.3 eV \cite{reininger83}) the large atom density of  $\sim$10$^{22}\, \mathrm{cm}^{-3}$ means that a cross-section as low as 35 $\mu$b is sufficient. So, although sub-threshold photoionization is very unlikely, the interaction probabilities required are also extremely small; in addition, the non-zero width of the scintillation emission will play a favourable role.  Alternatively, minute concentrations of other species with low ionization thresholds may be responsible. Of these there are many candidates, both organic and inorganic. In addition to neutral species, it is also possible that negative ions previously created by electron attachment can be photoionised during their long drift towards the anode, which is as slow as $\sim$0.7 cm/s for an O$_{2}^{-}$ ion in liquid xenon at 1 kV/cm \cite{schmidt}.  However, the concentration of O$_{2}^{-}$ ions accumulated from electron attachment during calibration and background runs seems insufficient to explain the rate observed.  It is also believed that these ions do not affect the electroluminescence field by more than 10\%.  Finally, we mention photoionization induced by the well-known `n$=$1' exciton in liquid xenon, which lies below the intrinsic threshold at 8.4 eV \cite{beaglehole,reshotko,hitachi_photo}. Although excitons do not cause photoionization directly, they can transfer the excitation to impurities onto which they become trapped and ionise them in a Penning-type interaction.  Clearly, further work is required to establish the most likely species undergoing photoionization in liquid xenon systems due to luminescence photons.

In summary, a population of small signals in the ZEPLIN-II low-background data were investigated and identified as single ionization electrons emitted into the gas region.  The electroluminescence yield shows good agreement with previous measurements for saturated xenon vapour, also displaying the expected pressure dependence.  The detection of single electrons shows the excellent sensitivity of two-phase xenon systems in the ionization channel, vital for high-sensitivity experiments such as WIMP dark matter searches.   The production of electrons occurs throughout the liquid xenon, with photoionization of contaminant species (either directly or exciton-induced) being the most probable production mechanism.  This may have possible implications for future experiments, as single electrons not directly related to particle interactions  in the xenon could constitute an additional background when searching for very small ionization signals.

The authors would like to acknowledge the work of the ZEPLIN-II collaboration and thank D. Akimov (ITEP, Moscow) for useful discussions on the subject of xenon physics.  This work was funded by UK Particle Physics and Astronomy Research Council (PPARC), the Portuguese Funda\c c\~ao para a Ci\^encia e a Tecnologia (project POCI/FP/FNU/63446/2005), the US Department of Energy (grant number DE-FG03-91ER40662) and the US National Science Foundation (grant number PHY-0139065).  

\end{document}